\documentclass{astrobul}
\usepackage{amsmath}
\usepackage{graphicx}
\usepackage{natbib}
\usepackage{color}
\begin{document}

\journalinfo{2026}{81}{2}{1}[12]

\title{START OF~ORBIT LIBRATIONS AND THE~BAR GROWTH TIMESCALE}

\author{E.~N.~Podzolkova\address{1,2}\email{podzolkova.en14@physics.msu.ru},
	A.~M.~Melnik\address{1},
	\addresstext{1}{Sternberg Astronomical Institute, Lomonosov Moscow
		State University, Universitetskij pr. 13, Moscow 119234, Russia}
	\addresstext{2}{Faculty of Physics, Lomonosov Moscow State University, Leninskie
		Gory 1-2, Moscow 119991, Russia} }
	
\shortauthor{PODZOLKOVA and MELNIK}

\shorttitle{START OF ORBIT LIBRATIONS}

\submitted{September 9, 2025; revised December 25, 2025; accepted February 26, 2026}

\begin{abstract}
	We study a dynamical model of the Galaxy with an analytical bar that reproduces the radial velocity $V_\mathrm{R}$ profiles as a function of the Galactocentric distance $R$ obtained from the Gaia\,DR3 data. The model radial velocity profiles show a periodic increase in $V_\mathrm{R}$ caused by orbits trapped into libration near the outer Lindblad resonance (OLR). To determine the moment when the librations start, we built a set of additional models differing only in the bar growth time $T_\mathrm{g}$. The temporal dependences of the radial velocity $V_\mathrm{R}(t)$ in the models with different $T_\mathrm{g}$ retain their shape but are shifted relative to each other in time $t$. The shift providing the best agreement between the model dependences is proportional to $T_\mathrm{g}$ with the coefficient $k = 0.54 \pm 0.02$. Orbit librations do not start when the bar reaches its full strength, but when it attains only 54\% of its maximum strength. Since the maximum bar strength in the models is $Q_\mathrm{b} = 0.314$, the librations start when the bar strength reaches $Q_\mathrm{b} = 0.170$.
	
\end{abstract}

\section{1. Introduction}

There is compelling evidence for the presence of a bar in the Galaxy \citep{Blitz1991, Dwek1995, Benjamin2005, CabreraLavers2007, Pohl2008, Gerhard2011, Li2012, NessLang2016}. Various estimates of the bar pattern velocity lie in the range of $\Omega_\mathrm{b} = 40$--60~km~s$^{-1}$~kpc$^{-1}$ \citep{Kalnajs1991, Dehnen2000, Minchev2007, Gerhard2011, Antoja2014, Bobylev2016, Melnik2019, Sanders2019, Melnik2021, Asano2022}. 

Rotation of the bar with the angular velocity $\Omega_\mathrm{b}$ induces resonances between the epicyclic frequency of stellar oscillations $\kappa$ and the orbital frequency relative to the bar, $\Omega - \Omega_\mathrm{b}$. The locations of the bar resonances in the galactic disk are determined from the condition
\begin{equation}
	\frac{m}{n} = \frac{\kappa}{\Omega - \Omega_\mathrm{b}},
\end{equation}
where $m$ is the number of full epicyclic oscillations during $n$ revolutions relative to the bar. Among these resonances, the two Lindblad resonances of the bar play a particularly important role: the outer Lindblad resonance (OLR) corresponding to the ratio $m/n = -2/1$, and the inner Lindblad resonance (ILR) corresponding to $m/n = 2/1$. Ultraharmonic resonances ($m/n = \pm 4/1$) also play an important role \citep{Contopoulos1983, Contopoulos1989, Athanassoula1992a}.

In many barred galaxies, elliptical resonance rings form in the vicinity of these resonances. Near the OLR of the bar, two types of outer rings are formed: $\mathrm{R}_1$ rings elongated perpendicular to the bar and located slightly closer to the galactic center, and $\mathrm{R}_2$ rings located farther from the center and elongated parallel to the bar \citep{Schwarz1981, Buta1991, Byrd1994, Buta1995, Buta1996, Rautiainen1999, Rautiainen2000}.

The strength of the non-axisymmetric perturbations introduced by the bar significantly affects resonant processes. For quantitative characterization, the parameter $q_t(R)$ is often used; it represents the ratio of the maximum tangential force to the azimuthally averaged radial force at the distance $R$ \citep{Sanders1980, Combes1981, Athanassoula1983, Block2001, Laurikainen2002, Buta2004, DiazGarcia2016}:
\begin{equation}
	q_t(R) = \frac{\max\,\,|F_\mathrm{T}(R)|}
	{\langle|F_\mathrm{R}(R)|\rangle}.
\end{equation}

The bar strength is defined as the maximum value of $q_t(R)$ over $R$ \citep{Buta2001}:
\begin{equation}
	Q_\mathrm{b} = \max\,\,q_t(R).
\end{equation}

\cite{Buta2005} showed that bar classification based solely on $Q_\mathrm{b}$ agrees well with visual classification. \cite{DiazGarcia2016} found that galaxies of visual types $\text{SB}$, $\text{SA}\underline{\text{B}}$, $\text{SAB}$, and $\text{S}\underline{\text{A}}\text{B}$ differ primarily in the bar strength. \cite{Lee2020} classified galaxies according to the distribution of $q_t$ in the galactic plane and showed that $Q_\mathrm{b} = 0.25$ can be used as a boundary between strong and weak bars. 

Strong bars are generally more elongated and have larger relative lengths (the ratio of the bar length to the photometric radius of the galaxy, $R_{25.5}$) than weak bars \citep{DiazGarcia2016}. \cite{Cuomo2019} also demonstrated that strong bars are longer than weak ones and have larger corotation radii $R_\mathrm{CR}$. Strong bars rotate more slowly: when $R_\mathrm{CR}/R_\mathrm{b} > 1.4$, the exchange of angular momentum with the halo and bulge becomes sufficiently efficient to slow down the bar \citep{Debattista2000, Lee2022}.

In galaxies with strong bars, outer resonance rings of type $\text{R}_2$ elongated parallel to the bar tend to dominate, whereas in galaxies with weak bars, $\text{R}_1$ rings elongated perpendicular to the bar are more common \citep{Comeron2014}. In models with strong bars, rings form more rapidly and are more clearly defined \citep{Schwarz1984}. Simulations show that bars tend to become stronger and thinner with time \citep{Athanassoula2003, Martinezvalpuesta2006}.

Barred galaxy models also produce stellar groups similar to the Hercules stream---a moving group of stars lagging behind the velocity of the rotation curve in the azimuthal direction and having large radial velocities directed away from the Galactic center \citep{Fux2001, Minchev2007, Antoja2014, Monari2017, Hunt2018}. \cite{Dehnen2000} showed that as the bar strength increases, the Hercules stream becomes more pronounced. The Hercules stream also emerges in our models \citep{Melnik2025}.

In this paper, we investigate the relation between the moment of the start of librations in the direction of orbital elongation and the bar growth timescale $T_\mathrm{g}$. We demonstrate that the start of librations in all models occurs at the same bar strength $Q_\mathrm{b}$ and does not depend on the bar growth time $T_\mathrm{g}$.

\section{2. Models}

We use a 2D model of the Galaxy including a bar, bulge, exponential disk, and spherical halo. The exponential disk has the mass $M_\mathrm{d} = 3.25 \times 10^{10}\,M_\odot$ and the characteristic scale length $R_\mathrm{d} = 2.5~\text{kpc}$. The bulge is represented by a Plummer sphere with the mass $M_\mathrm{bg} = 5 \times 10^9\,M_\odot$ \citep{Nataf2017, Fujii2019}. The halo is modeled as an isothermal sphere \citep{Binney2008}. 

The bar is represented by a Ferrers ellipsoid with the mass $M_\mathrm{b} = 1.2 \times 10^{10}~M_\odot$ and the semi-axes $a = 3.5$ and $b = 1.35$~kpc \citep{deVaucouleurs1972, Athanassoula1983, Pfenniger1984, Sellwood1993, Binney2008, Fujii2019}. The bar pattern velocity is $\Omega_\mathrm{b} = 55~\text{km}~\text{s}^{-1}~\text{kpc}^{-1}$. The corotation radius (CR) of the bar is located at $R_\mathrm{CR} = 4.04$~kpc. The outer Lindblad resonance is located at $R_\mathrm{OLR} = 7.00$~kpc.

\begin{figure}
		\centering
	\includegraphics[width=0.45\textwidth]{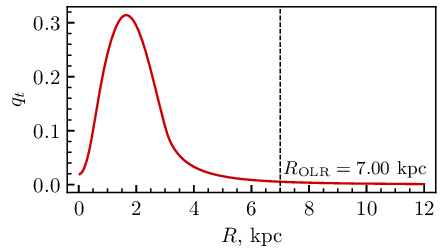}
	\caption{Dependence of $q_t$ on the Galactocentric distance $R$ after the bar reaches its full strength ($T_\mathrm{g} = 0.45$~Gyr). The maximum value $q_t = 0.3142$ is reached at $R = 1.64$~kpc.}
	\label{fig:qt(r)}
\end{figure}

At the initial time instant, the bar potential is axisymmetric, and the non-axisymmetric perturbations grow linearly over four bar rotations, corresponding to the growth time $T_\mathrm{g} = 0.45$~Gyr \citep{Rautiainen2000, Rautiainen2010}:
\begin{equation}
	\begin{cases}
		Q_\mathrm{b}^*(t) = Q_\mathrm{b}\;\cfrac{t}{T_\mathrm{g}}, & \text{for\,\,\,} 0 \le t<T_\mathrm{g}, \\
		Q_\mathrm{b}^*(t) = Q_\mathrm{b}, 
		& \text{for\,\,\,} t \ge T_\mathrm{g},
	\end{cases}
	\label{eq:growth}
\end{equation}
\noindent
where $Q_\mathrm{b} = 0.3142$ corresponds to the distance $R = 1.64$~kpc from the Galactic center (Figs.~\ref{fig:qt(r)} and \ref{fig:qb(t)}).

The position angle of the Sun with respect to the bar is adopted as $\theta_\odot = -45^\circ$. Since the model has the order of symmetry $m = 2$, angles of $-45^\circ$ and $135^\circ$ are equivalent. The model includes $N = 2\times{10^6}$ massless particles. For more details, see \cite{Melnik2021}.

We built a set of additional models that differ from the reference model with $T_\mathrm{g} = 0.45$~Gyr only in the bar growth timescale: $T_\mathrm{g} = 1.0,\,1.5,\,2.0,\,2.5$, and $3.0$~Gyr. For models with $T_\mathrm{g} = 1.0$ and $1.5$~Gyr, the simulation time was $7.5$~Gyr, while for all other models, it was $8.0$~Gyr. This ensures that each model includes at least $5.0$~Gyr of evolution after the bar has fully grown. The simulation time of the reference model with $T_\mathrm{g} = 0.45$~Gyr was also extended to $7.0$~Gyr.

We built profiles of the median radial ($V_\mathrm{R}$) and azimuthal ($V_\mathrm{T}$) velocity components of stars in the model disk. The median velocities $V_\mathrm{R}$ and $V_\mathrm{T}$ were calculated within the sector \mbox{$|\theta-\theta_\odot| < 15^\circ$} and in radial bins of width $\Delta R = 250$~pc. The model velocity profiles were averaged over time intervals of $0.5$~Gyr.

\section{3. Observations}
We selected stars from the Gaia\,DR3 catalog \citep{Prusti2016, Katz2018, Brown2021, Vallenari2023} with reliable parallaxes ($\varpi/\epsilon_{\varpi} > 5$), re-normalized unit weight error $\text{RUWE} < 1.4$ \citep{Lindegren2018}, and measured line-of-sight velocities $V_\mathrm{r}$. From this sample, we selected stars located within the sector of the Galactocentric angles $|\theta| < 15^\circ$ and close to the Galactic plane, $|z| < 200$~pc. The final sample includes $9.7 \times 10^6$ stars.

The radial velocity dispersion of the sample at the solar radius is $\sigma_\mathrm{R} = 32.0$~km~s$^{-1}$, which is characteristic of the old thin-disk population. According to the Gaia~FLAME data, in the region of primary interest for our study ($|z| < 200$~pc, $|\theta| < 15^\circ$, $6 \le R \le 9$~kpc), 73\% of stars are older than 2~Gyr. This is consistent with estimates such as those of \cite{Yu2018}, who have derived ages greater than 2.3~Gyr for the radial velocity dispersion $\sigma_\mathrm{R} \sim 30$~km~s$^{-1}$.

The Galactocentric distance of the Sun is adopted to be $R_0 = 7.5$~kpc \citep{Glushkova1998, Nikiforov2004, Eisenhauer2005, Bica2006, Nishiyama2006, Feast2008, Groenewegen2008, Reid2009, Dambis2013, Francis2014, Boehle2016, Branham2017, Iwanek2023}. In general, adopting a distance in the range of 7--9~kpc has little effect on our results. The main result of this study---the start time of orbital librations---is obtained from a comparison of different models and does not depend on the specific choice of $R_0$. Changing $R_0$ affects only the absolute positions of resonances, but does not alter the distance between the Sun and the OLR radius. For example, adopting $R_0 = 8.1$~kpc \citep{Bobylev2021} places the OLR at $R_\mathrm{OLR} = 7.6$~kpc.

We also built observational profiles of the velocity components $V_\mathrm{R}$ and $V_\mathrm{T}$. The median values of $V_\mathrm{R}$ and $V_\mathrm{T}$ as functions of the Galactocentric distance $R$ were calculated in radial bins of width $\Delta R = 250$~pc.

\section{4. FORMATION OF HUMPS}

\begin{figure*} 
	\centering
	\includegraphics[width=0.9\textwidth]{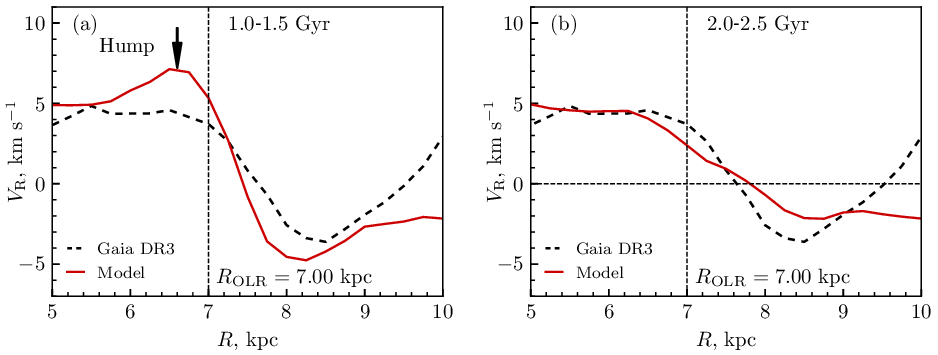}  
	\caption{Radial velocity $V_\mathrm{R}$ profiles derived for the Gaia\,DR3 stars (black dashed curve) and for the stars of the model with $T_\mathrm{g} = 0.45$~Gyr (red curve) averaged over different time intervals. (a)~Time interval $t = 1.0\text{--}1.5$~Gyr. A hump forms at $R = 6\text{--}7$~kpc and is indicated by the arrow. (b)~Time interval $t = 2.0\text{--}2.5$~Gyr. The hump is absent.}
	\label{fig:humps}
\end{figure*}

We discovered the formation of humps in the radial velocity profiles $V_\mathrm{R}(R)$. Figure~\ref{fig:humps} shows the radial component of the median velocity $V_\mathrm{R}$ as a function of the Galactocentric distance $R$ for the Gaia~DR3 stars (black dashed curve) and for the stars of the model disk (red curve) averaged over the time intervals (a) $t = 1.0\text{--}1.5$~Gyr and (b) $t = 2.0\text{--}2.5$~Gyr from the start of the simulation. In Fig.~\ref{fig:humps}a, the model profile shows an increase in $V_\mathrm{R}$ in the range $R = 6\text{--}7$~kpc (a hump), whereas in Fig.~\ref{fig:humps}b the hump is absent.

\cite{Weinberg1994} showed that orbits trapped by the OLR can change their orientation with respect to the bar major axis in two ways: either the direction of orbital elongation precesses continuously without any angular restriction, or it varies within a limited angular range.

We have shown that the formation of humps is periodic. The humps are supported by orbits trapped into librations near the OLR of the bar. The direction of elongation of these orbits librates with respect to the bar major axis. Such orbits contribute additional negative radial velocities ($V_\mathrm{R} < 0$) to the region of the humps formation: $|\theta - \theta_\odot| < 15^\circ$ and $R = 6\text{--}7$~kpc, but when stars collectively leave this region, humps are formed \citep{Melnik2024}.

\begin{figure*}[t]
	\centering
	\includegraphics[width=0.75\textwidth]{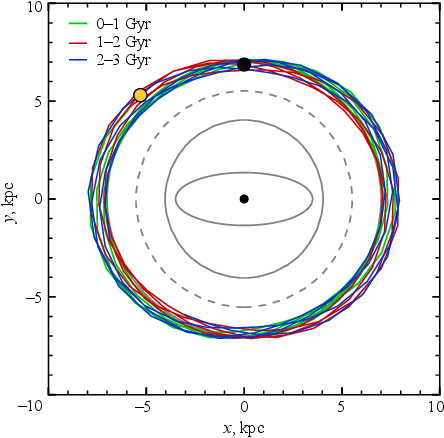} 
	\caption{Example of an orbit supporting the humps. The Galaxy rotates counterclockwise. The orbit is shown in the reference frame of the rotating bar, in which the star moves clockwise. The bar is shown as the gray ellipse. Positions of CR and OLR are shown by solid gray lines, and the $-4/1$ resonance by the dashed gray line. The initial position of the star is marked by the black circle, and the assumed position of the Sun relative to the bar is marked by the yellow circle. The initial conditions are $R(0) = R_\mathrm{OLR} + 0.1$~kpc, $\theta(0) = 90^\circ$, $V_\mathrm{R}(0) = 0$, and $V_\mathrm{T}(0) = V_\mathrm{c}$. Orbit segments corresponding to the intervals $0\text{--}1$, $1\text{--}2$, and $2\text{--}3$~Gyr are shown in green, red, and blue, respectively. The direction of orbital elongation librates: during $0\text{--}1$ and $2\text{--}3$~Gyr (green and blue curves), the orbit is tilted to the right (opposite to the direction of the Galactic rotation), whereas during $1\text{--}2$~Gyr (red curve) it is tilted to the left.}
	\label{fig:orb}
\end{figure*}

Most orbits responsible for the humps support the outer ring $\text{R}_2$. A typical orbit of such a star is shown in Fig.~\ref{fig:orb}. The Galaxy rotates counterclockwise; however, in the reference frame of the rotating bar, the considered star located beyond the corotation radius ($R > R_\mathrm{CR}$) rotates clockwise. At the initial time, the star is located near the OLR ($R(0) = R_\mathrm{OLR} + 0.1$~kpc) on the bar minor axis ($\theta(0) = 90^\circ$) with the velocities $V_\mathrm{R}(0) = 0$ and $V_\mathrm{T}(0) = V_\mathrm{c}$, where $V_\mathrm{c} = 225$~km~s$^{-1}$ is the velocity of the rotation curve at the solar radius.

Orbit segments corresponding to the time intervals $0\text{--}1$, $1\text{--}2$, and $2\text{--}3$~Gyr from the beginning of the simulation are shown in green, red, and blue, respectively. The orbit is generally elongated parallel to the bar, thus supporting the $\text{R}_2$ ring. During the intervals $0\text{--}1$ and $2\text{--}3$~Gyr, the orbit is tilted to the right---opposite to the direction of Galactic rotation---whereas during $1\text{--}2$~Gyr it is tilted to the left and supports hump formation.

\begin{figure*}
	\centering
	\includegraphics[width=0.75\textwidth]{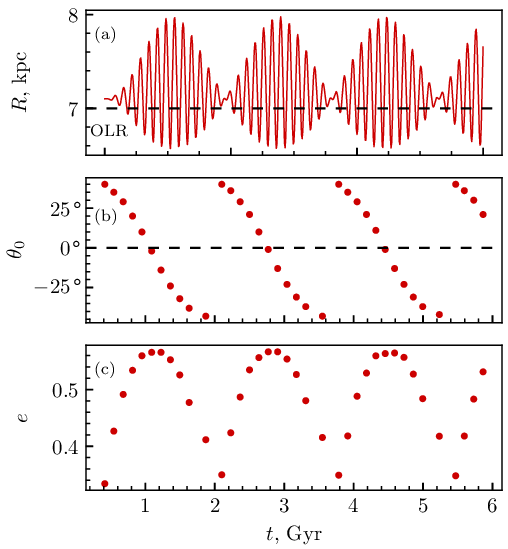}
	\caption{(a)~Variation of the Galactocentric distance $R$ with time for the star shown in Fig.~\ref{fig:orb}. The oscillations clearly show a beat pattern. The horizontal dashed line indicates the OLR radius. (b)~Time evolution of the orbit elongation angle $\theta_0$ computed over one radial oscillation period. The angle decreases gradually from $+40^\circ$ to $-40^\circ$ and then rapidly returns to its initial value. (c)~Time variation of the orbital eccentricity $e$.} 	
	\label{fig:orb_par}
\end{figure*}

Figure~\ref{fig:orb_par}a shows the variation of the Galactocentric distance $R$ with time. The oscillations of $R$ display a beat pattern. The beats arise between the epicyclic frequency $\kappa$ and the frequency with which the star encounters the bar perturbation, $2(\overline{\dot{\theta}} - \Omega_\mathrm{b})$:
\begin{equation}
	\omega_\mathrm{bt} = \kappa(\overline{R}) + 2(\overline{\dot{\theta}} - \Omega_\mathrm{b}).
\end{equation} 

The beats in $R$ shown in Fig.~\ref{fig:orb_par}a have a period of $P = 1.71 \pm 0.03$~Gyr. This period coincides with periods of the $\theta_0$ variation and the orbital eccentricity.

Figure~\ref{fig:orb_par}b shows the evolution of the orbital elongation angle $\theta_0$ defined over one radial oscillation period (from one crossing of the mean orbital radius with negative radial velocity, $V_\mathrm{R} < 0$, to the next). The angle~$\theta_0$ is measured from the bar major axis in the direction of the Galactic rotation (counterclockwise). The angle $\theta_0$ gradually decreases from $\theta_0 = +40^\circ$ to $-40^\circ$, and then rapidly returns to its initial value.

The orbital eccentricity $e$ shown in Fig.~\ref{fig:orb_par}c varies within the range of $0.33\text{--}0.57$.
    
An analysis of the stellar distribution in the model disk reveals periodic enhancement of either leading or trailing segments of the outer resonance rings, with a period $P \approx 2$~Gyr. It turns out that the morphological variations of the outer rings, as well as the formation of humps, are supported by librating orbits \citep{Melnik2023}.

\section{5. RESULTS}
\subsection{5.1 Time Shift of the $V_\mathrm{R}$ Oscillations}
We find that the librations of the orbital elongation direction do not start at the beginning of the simulation, but only after the bar reaches a sufficient strength to trigger them.

To determine the moment of start of the librations, we analyzed the radial velocity distributions $V_\mathrm{R}$ in models that differ only in the bar growth time.

\begin{figure*}
	\centering
	\includegraphics[width=0.75\textwidth]{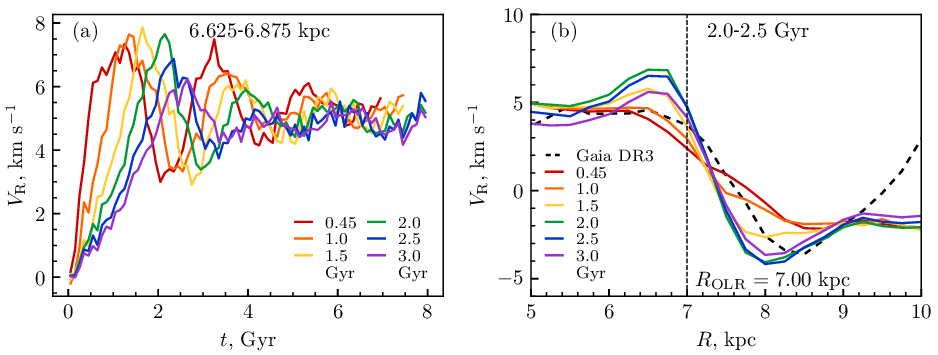}
	\caption{(a)~Radial velocity $V_\mathrm{R}$ as a function of time from the beginning of the simulation $t$ for models with different bar growth timescales $T_\mathrm{g}$ calculated in the radial bin $R = 6.625\text{--}6.875$~kpc. The curves have similar shapes but are shifted to later times as $T_\mathrm{g}$ increases. Periodic variations of $V_\mathrm{R}$ begin earlier for smaller $T_\mathrm{g}$. In addition to the shift, the oscillation amplitude decreases with increasing $T_\mathrm{g}$. (b)~Radial velocity profiles $V_\mathrm{R}(R)$ for models with different $T_\mathrm{g}$ (colored curves) and for Gaia~DR3 stars (black dashed curve). Each model profile is averaged over the interval $t = 2.0\text{--}2.5$~Gyr from the beginning of the simulation. The hump height varies between models: as $T_\mathrm{g}$ increases, the hump height first increases and then decreases.}   
	\label{fig:profs_before}  
\end{figure*}

Figure~\ref{fig:profs_before}a shows $V_\mathrm{R}$ as a function of time $t$ in the radial bin $R = 6.625\text{--}6.875$~kpc for the reference model with $T_\mathrm{g} = 0.45$~Gyr (red curve) and for the additional models (colored curves). This bin has been chosen because the hump height in the reference model reaches its maximum there. As $T_\mathrm{g}$ increases, the curves shift systematically to later times while preserving their overall shape. Note also that the amplitude of the $V_\mathrm{R}$ oscillations decreases with increasing $T_\mathrm{g}$.

Figure~\ref{fig:profs_before}b shows the radial velocity profiles $V_\mathrm{R}(R)$ for models with different $T_\mathrm{g}$, averaged over $t = 2.0\text{--}2.5$~Gyr. With increasing $T_\mathrm{g}$, the profiles first show an increase of $V_\mathrm{R}$ at \mbox{$R = 6\text{--}7$}~kpc and then a decrease, i.e., they trace consecutive stages of hump formation and disappearance. Therefore, the same averaging interval $t = 2.0\text{--}2.5$~Gyr corresponds to different phases of the oscillations in different models. The observational profile $V_\mathrm{R}(R)$ derived from Gaia~DR3 is also shown (black dashed curve).

The relative shifts between the curves in Fig.~\ref{fig:profs_before}a indicate that periodic librations of the orbital elongation direction begin not when the bar has fully grown, but at an earlier time when the bar reaches a threshold strength $Q_\mathrm{b}$ sufficient to trigger the librations. To determine this threshold, the curves in Fig.~\ref{fig:profs_before}a must be shifted in time so as to bring them into the best agreement. We introduced a time-shift coefficient $k$ common to all models. For each model, the time from the start of the simulation, $t$, is shifted by an amount proportional to $T_\mathrm{g}$:
\begin{equation}
	t^{\prime} = t - k T_\mathrm{g}.    
\end{equation} 

To find the optimal shift coefficient $k_0$ that yields the best match between the $V_\mathrm{R}(t^{\prime})$ dependences of all models, we minimized the sum of $\chi^2$ functionals:
\begin{equation}    
	\chi^2(k) = \sum_{i<j}\sum_{n}\frac{\Delta V_{ij}^2(t^{\prime}_n, k)}{\sigma^2_{in} + \sigma_{jn}^2},
\end{equation}
\noindent
where
\begin{equation}
	\Delta V_{ij}(t^{\prime}_n,~k) = V_{{R},\,i}(t^{\prime}_n) - V_{{R},\,j}(t^{\prime}_n).
\end{equation}

The summation was performed over 15 different pairs of curves $\left(i,\,\,j\right)$ and 54~shifted-time points $t^{\prime}_n$ in the range of $-400 \le t^{\prime}_n < 5000$~Myr with a step of 100~Myr. This set of shifted-time points is present in all models for any shift coefficient within the range of $k = 0\text{--}1$. The uncertainties of the median radial velocities at each time point were assumed to be $\sigma = 0.62$~km~s$^{-1}$ \citep{Melnik2021}.

\begin{figure}
	\centering
	\includegraphics[width=0.45\textwidth]{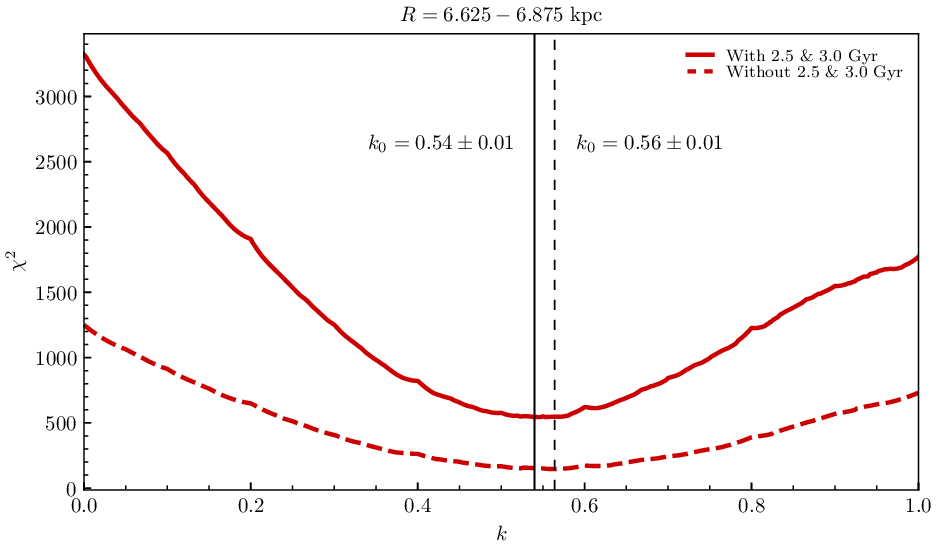} 
	\caption{$\chi^2$ as a function of the shift coefficient $k$ in the radial bin $R = 6.625\text{--}6.875$~kpc. The minima of $\chi^2$ are marked by the vertical lines. For the set excluding $T_\mathrm{g} = 2.5$ and $3.0$~Gyr (dashed curve), the minimum corresponds to $k_0 = 0.56\pm{0.01}$, while for the set including all models (solid curve) it corresponds to $k_0 = 0.54\pm{0.01}$.}  
	\label{fig:chi}
\end{figure}

Figure~\ref{fig:chi} shows $\chi^2$ as a function of the shift coefficient~$k$ in the radial bin $R = 6.625\text{--}6.875$~kpc for two model sets: one including all models (solid curve) and one excluding the models with $T_\mathrm{g} = 2.5$ and $3.0$~Gyr (dashed curve). For the latter two models, the $V_\mathrm{R}(t)$ dependence has a noticeably smaller oscillation amplitude than in the others. For the full set, the minimum $\chi^2_\mathrm{min}$ corresponds to $k_0 = 0.54 \pm 0.01$, while for the set excluding $T_\mathrm{g} = 2.5$ and $3.0$~Gyr, it corresponds to $k_0 = 0.56 \pm 0.01$. Further we consider the full set including all models.

\begin{figure*}
	\centering 
	\includegraphics[width=0.75\textwidth]{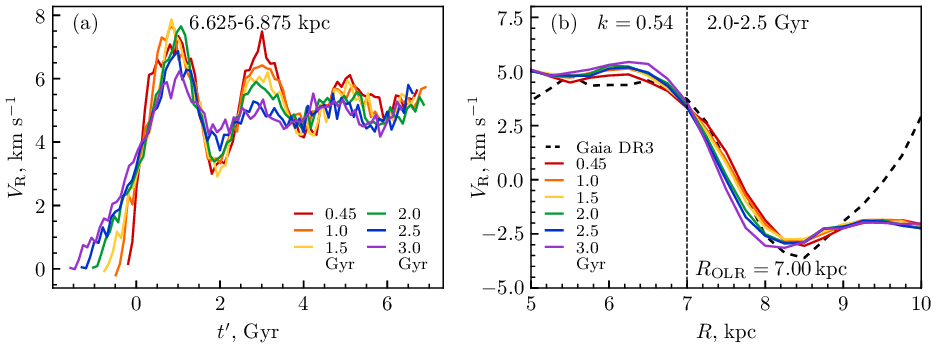}
	\caption{(a)~Radial velocity $V_\mathrm{R}$ as a function of the shifted time $t^{\prime} = t - k_0 T_\mathrm{g}$ in the radial bin $R = 6.625\text{--}6.875$~kpc (cf.~Fig.~\ref{fig:profs_before}a). The orbital librations begin simultaneously in all models at $t^{\prime} = 0$~Gyr. The oscillation amplitude decreases with increasing $T_\mathrm{g}$. (b)~Radial velocity profiles $V_\mathrm{R}(R)$ averaged over the shifted-time interval $t^{\prime} = 2.0\text{--}2.5$~Gyr (cf.~Fig.~\ref{fig:profs_before}b). After the shift, none of the models shows pronounced humps; the profiles for all $T_\mathrm{g}$ become nearly identical.}
	\label{fig:profs_after}
\end{figure*}

After applying the shift $t^{\prime} = t - k_0 T_\mathrm{g}$ \mbox{($k_0 = 0.54$)}, the orbital elongation directions librate nearly synchronously starting at $t^{\prime} = 0$. Figure~\ref{fig:profs_after}a, analogous to Fig.~\ref{fig:profs_before}a, shows the result of the time-shift procedure. The $V_\mathrm{R}(t^{\prime})$ curves in the radial bin $R = 6.625\text{--}6.875$~kpc agree well and show the same oscillation period $P \approx 2$~Gyr starting at $t^{\prime} = 0$. The oscillation amplitude decreases with increasing $T_\mathrm{g}$ and drops sharply for $T_\mathrm{g} > 2.0$~Gyr. This may be related to the fact that when $T_\mathrm{g}$ exceeds the characteristic period $P \approx 2$~Gyr, the phase coherence of the oscillations is lost more rapidly.

Figure~\ref{fig:profs_after}b, analogous to Fig.~\ref{fig:profs_before}b, shows that after the time shift the $V_\mathrm{R}(R)$ profiles no longer display pronounced humps. To first order, the profiles for all models can be regarded as similar in shape. If a different shifted-time interval were chosen for averaging, all profiles would show the same hump corresponding to a different stage of the hump formation.

\begin{figure}
	\centering
	\includegraphics[width=0.45\textwidth]{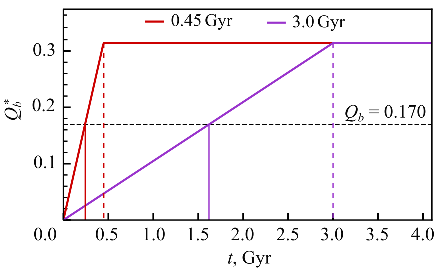}
	\caption{Schematic illustration of the bar strength growth (Equation~\ref{eq:growth}) for two models: $T_\mathrm{g} = 0.45$~Gyr (solid red line) and $T_\mathrm{g} = 3.0$~Gyr (solid purple line). The bar strength increases linearly to the maximum value $Q_\mathrm{b} = 0.3142$ over the time interval $T_\mathrm{g}$ (vertical dashed lines). The horizontal black dashed line indicates $Q_\mathrm{b} = 0.1696$ at which the librations of the orbital elongation direction begin. The solid vertical lines mark the times from the start of the simulation at which the librations begin.}
	\label{fig:qb(t)}
\end{figure}

Since the bar strength in our models grows linearly (Equation~\ref{eq:growth}), the value $k_0 = 0.54\pm{0.01}$ implies that periodic librations of the orbital elongation direction start when the bar reaches $54\pm1\%$ of its maximum strength $Q_\mathrm{b} = 0.3142$. Therefore, the librations begin when the bar reaches $Q_\mathrm{b} = 0.1696\pm{0.0025}$. This value is only slightly below $Q_\mathrm{b} = 0.25$, which is commonly used as an approximate boundary between weak and strong bars. Figure~\ref{fig:qb(t)} illustrates the time evolution of the bar strength for two models: with $T_\mathrm{g} = 0.45$~Gyr and $T_\mathrm{g} = 3.0$~Gyr. The bar strength grows linearly and reaches its maximum $Q_\mathrm{b} = 0.3142$ at $t = T_\mathrm{g}$. The horizontal black dashed line marks $Q_\mathrm{b} = 0.1696$ at which the librations of the orbital elongation direction begin. 

\subsection{5.2 Fitting the $V_\mathrm{R}$ Oscillations}

\begin{figure*} 
	\centering
	\includegraphics[width=0.75\textwidth]{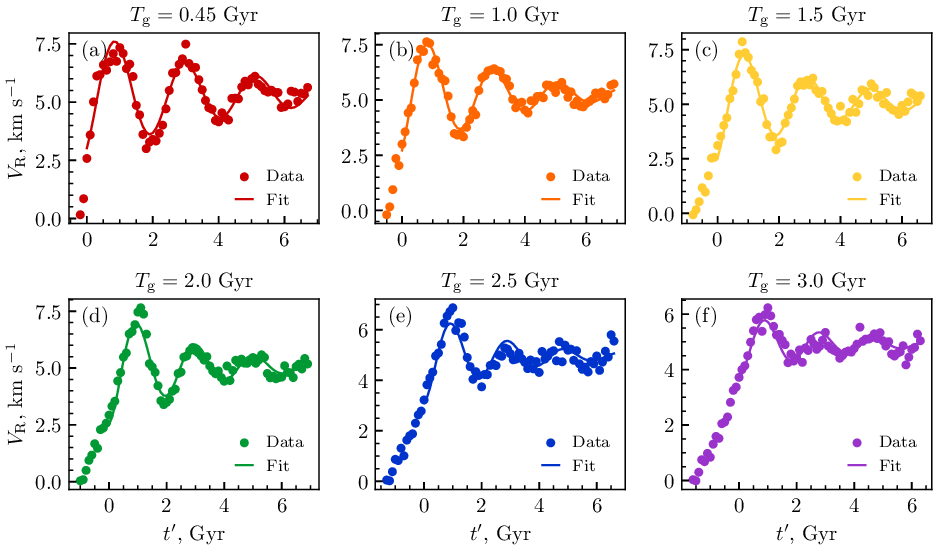}
	\caption{Radial velocity $V_\mathrm{R}$ as a function of the shifted time $t^{\prime}$ in different models (colored points) and fits with damped oscillations (Equation~\ref{eq:fit}, colored curves). The bar growth time in each model is indicated above the corresponding panel. The fits were performed from the start of the orbital librations ($t^{\prime} = 0$). The initial amplitude decreases with increasing $T_\mathrm{g}$ (see Fig.~\ref{fig:fit_params}a).}
	\label{fig:fits}
\end{figure*}

We fitted the dependences of the radial velocity $V_\mathrm{R}$ on the shifted time $t^{\prime}$ for different values of $T_\mathrm{g}$ (Fig.~\ref{fig:profs_after}a) with damped harmonic oscillations:

\begin{equation}    
	V_\mathrm{R}(t^{\prime}) = A e^{-\lambda t^{\prime}} \sin\left(\cfrac{2\pi}{P} t^{\prime} + \phi \right) + C,   
	\label{eq:fit}   
\end{equation} 
where $A$ is the initial amplitude, $\lambda$ is the damping coefficient, $P$ and $\phi$ are the period and initial phase of the oscillations, and $C$ is the mean value of $V_\mathrm{R}$. Figure~\ref{fig:fits} shows the fitting results. The fits were performed from the start of the orbital librations, i.e., from $t^{\prime} = 0$. Note the good agreement between the data and the fits.

\begin{figure*}
	\centering
	\includegraphics[width=0.75\textwidth]{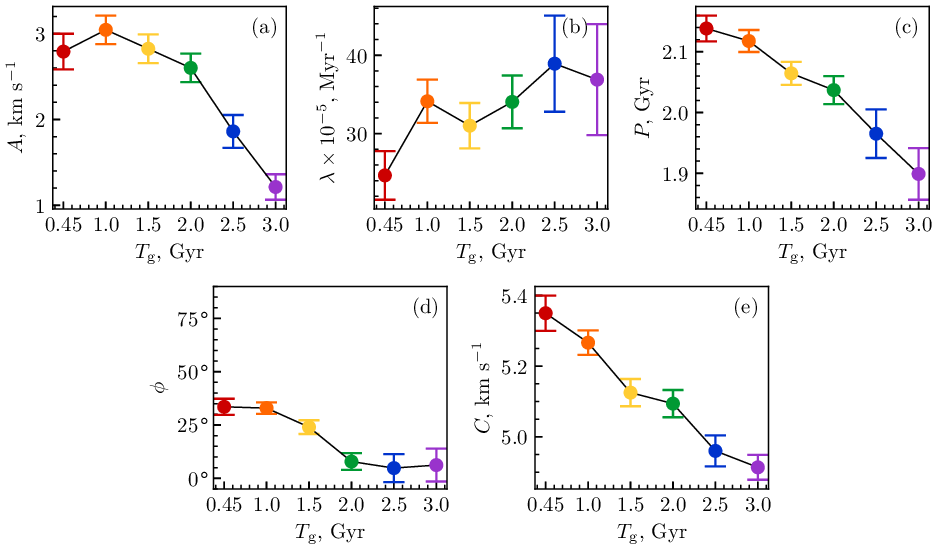}
	\caption{Fit parameters obtained by fitting $V_\mathrm{R}(t^{\prime})$ with damped harmonic oscillations (Equation~\ref{eq:fit}) for different bar growth times $T_\mathrm{g}$: initial amplitude $A$ (a), damping coefficient $\lambda$ (b), period $P$ (c), initial phase $\phi$ (d), mean velocity $C$ (e).}    
	\label{fig:fit_params} 
\end{figure*}

Figure~\ref{fig:fit_params} shows the dependence of the fit parameters (Equation~\ref{eq:fit}) on the bar growth time. The initial amplitude $A$ (Fig.~\ref{fig:fit_params}a) reaches a maximum at $T_\mathrm{g} = 1.0$~Gyr and drops sharply for $T_\mathrm{g} > 2.0$~Gyr. Overall, $A$ decreases by about 60\%. The damping coefficient $\lambda$ increases by 37\% with increasing $T_\mathrm{g}$ (Fig.~\ref{fig:fit_params}b). The oscillation period $P$ (Fig.~\ref{fig:fit_params}c) decreases nearly linearly by 11\% from $P = 2.1$~Gyr to $P = 1.9$~Gyr as $T_\mathrm{g}$ increases. The initial phase $\phi$ is shown in Fig.~\ref{fig:fit_params}d. The value $T_\mathrm{g} = 1.5$~Gyr divides the models into two groups, $T_\mathrm{g} > 1.5$~Gyr and $T_\mathrm{g} < 1.5$~Gyr, within which the initial phases agree within $\sigma$, i.e., the orbital librations are nearly synchronous within each group. Overall, $\phi$ varies in the range of $5^\circ$--$34^\circ$. The mean velocity $C$ (Fig.~\ref{fig:fit_params}e) decreases nearly linearly by about 8\% with increasing $T_\mathrm{g}$.

\subsection{5.3 Impact of Additional Factors on $k_0$}

\begin{figure} 
	\centering
	\includegraphics[width=0.45\textwidth]{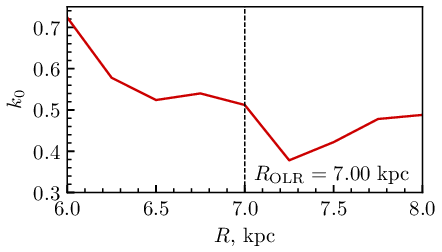} 
	\caption{Shift coefficient $k_0$ corresponding to the minimum $\chi^2$ as a function of the radial bin $R$. The values of $k_0$ vary between 0.38 and 0.72 for $R$ in the range of $6\text{--}8$~kpc; however, in the hump-formation region ($R = 6.5\text{--}7.0$~kpc), $k_0$ is nearly constant at $k_0 = 0.52\pm{0.02}$. A pronounced minimum is also visible at $R = 7.0\text{--}7.5$~kpc.}
	\label{fig:k0(R)}
\end{figure}

So far, we have considered only the radial bin in which the hump height reaches its maximum ($R = 6.625\text{--}6.875$~kpc), i.e., the bin where the effect of the orbital librations is most pronounced. We now consider other bins in the vicinity of the OLR. Figure~\ref{fig:k0(R)} shows the dependence of the shift coefficient $k_0$ (corresponding to the minimum $\chi^2$) on the choice of the radial bin over the range $R = 6\text{--}8$~kpc. The values of $k_0$ vary between 0.38 and 0.72. Overall, the variation of $k_0$ with bin choice indicates that the $V_\mathrm{R}$ profiles for different $T_\mathrm{g}$ are only approximately similar. In the region where humps form ($R = 6.5\text{--}7.0$~kpc), $k_0$ is nearly constant, $k_0 = 0.52\pm{0.02}$; therefore, a more realistic estimate of the uncertainty of $k_0$ is $\pm 0.02$. A pronounced minimum is seen at $R = 7.0\text{--}7.5$~kpc; we cannot explain its presence in terms of the formation and disappearance of the humps. The value of $k_0$ in bins where humps do not form is determined not by librating orbits but by other factors.

\begin{figure*}
	\centering
	\includegraphics[width=0.7\textwidth]{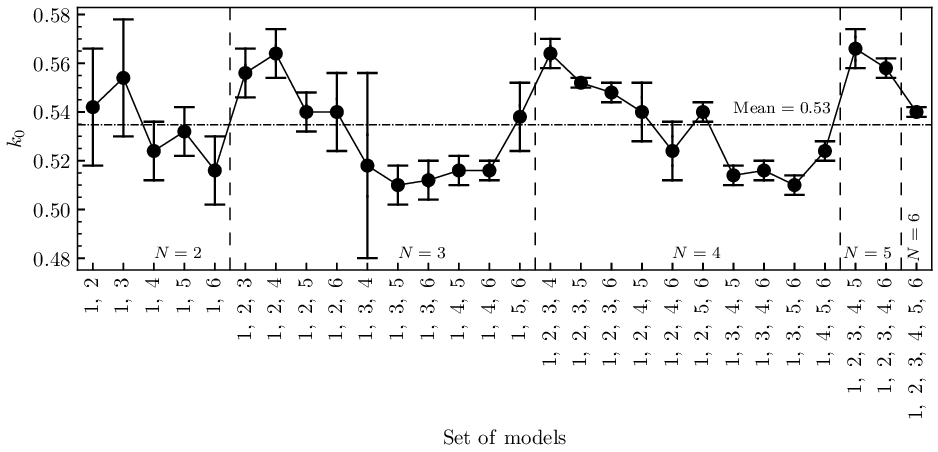}
	\caption{Shift coefficient $k_0$ corresponding to the minimum $\chi^2$ for different model sets. To avoid overly dense tick labels on the horizontal axis, the models are numbered from 1 to 6 in order of increasing $T_\mathrm{g}$. Only sets that include the reference model with $T_\mathrm{g} = 0.45$~Gyr are shown. The vertical dashed lines separate sets with different numbers of models, from $N = 2$ to $N = 6$. For sets containing the same number of models, $k_0$ decreases as the difference between the largest and smallest $T_\mathrm{g}$ values in the set increases. The mean value over all shown sets is $k_0 = 0.53$.}
	\label{fig:k0(set)}
\end{figure*}

As shown in Fig.~\ref{fig:chi}, excluding the models with $T_\mathrm{g} = 2.5$ and $3.0$~Gyr has little effect on the results. We~now consider other model sets that include the reference model with $T_\mathrm{g} = 0.45$~Gyr. Figure~\ref{fig:k0(set)} shows the shift coefficient $k_0$ (corresponding to the minimum $\chi^2$) for different model sets. The sets are grouped by the number of models they include (from $N = 2$ to $N = 6$), and sets with different $N$ are separated by the vertical dashed lines. For compactness, the tick labels on the horizontal axis do not list the $T_\mathrm{g}$ values; instead, the models are numbered from 1 to 6 in order of increasing $T_\mathrm{g}$. Among sets containing the same number of models, larger $k_0$ values are obtained for those sets in which the difference between the minimum and maximum $T_\mathrm{g}$ is smaller. The mean value of $k_0$ over all shown sets is $k_0 = 0.53 \pm 0.02$, close to our adopted value $k_0 = 0.54\pm{0.02}$. Depending on the model set, $k_0$ varies in the range of 0.51--0.57.

\section{6. CONCLUSIONS}
Using Galaxy models with an analytical bar, we investigated oscillations of the radial velocity $V_\mathrm{R}$ with time. The $V_\mathrm{R}$ oscillations are caused by orbits librating with respect to the bar major axis with a period of $\sim 2$~Gyr. We find that these orbital librations begin only after the bar reaches a certain strength.

To determine the moment of start of the librations of the orbital elongation direction, we built five additional models that differ from the reference model with $T_\mathrm{g} = 0.45$~Gyr only in the bar growth time: $T_\mathrm{g} = 1.0,\,1.5,\,2.0,\,2.5,$ and $3.0$~Gyr.

For each model, we computed $V_\mathrm{R}(t)$ in the radial bin $R = 6.625\text{--}6.875$~kpc. The $V_\mathrm{R}(t)$ curves for different $T_\mathrm{g}$ have similar shapes but are shifted relative to one another (Fig.~\ref{fig:profs_before}a). The radial velocity profiles $V_\mathrm{R}(R)$ built for different models trace successive stages of hump formation and disappearance (Fig.~\ref{fig:profs_before}b).

We determined the time-shift coefficient $k_0$ that provides the best match between the $V_\mathrm{R}(t)$ curves. The obtained value $k_0 = 0.54\pm0.02$ implies that the orbital librations start when the bar reaches $54\pm2\%$ of its maximum strength, which is $Q_\mathrm{b} = 0.3142$ in all models. Thus, the librations of the orbital elongation direction begin at $Q_\mathrm{b} = 0.170\pm0.006$ and do not depend on the bar growth timescale $T_\mathrm{g}$ (Fig.~\ref{fig:profs_after}a).

We fitted the dependences $V_\mathrm{R}(t^{\prime})$ for models with different bar growth times $T_\mathrm{g}$ (Fig.~\ref{fig:fits}) with damped oscillations (Equation~\ref{eq:fit}). Overall, the initial amplitude $A$ decreases with increasing $T_\mathrm{g}$. The decrease in the oscillation amplitude with increasing $T_\mathrm{g}$ may be related to a loss of coherence of the organized orbital librations with respect to the bar major axis as $T_\mathrm{g}$ increases. The oscillation period $P$ and the mean velocity $C$ decrease nearly linearly with increasing $T_\mathrm{g}$ by about 11\% and 8\%, respectively, while the damping coefficient $\lambda$ increases with $T_\mathrm{g}$. The value $T_\mathrm{g} = 1.5$~Gyr separates the models into two groups, $T_\mathrm{g} < 1.5$~Gyr and $T_\mathrm{g} > 1.5$~Gyr, within which the fitted phase $\phi$ is consistent within the uncertainties (Fig.~\ref{fig:fit_params}).

We examined the impact of additional factors on the shift coefficient $k_0$. In other radial bins, $k_0$ varies over the range of $0.38\text{--}0.72$, but changes only slightly within the hump-formation region ($R = 6.5\text{--}7.0$~kpc; Fig.~\ref{fig:k0(R)}). In the bins where humps do not form, the value of $k_0$ is determined not by librating orbits but by other factors.

Considering different model sets shows that for the sets containing the same number of models, $k_0$ decreases as the difference between the maximum and minimum $T_\mathrm{g}$ values in the set increases. Overall, depending on the chosen model set, the shift coefficient varies within $0.51\text{--}0.57$ (Fig.~\ref{fig:k0(set)}).

\section*{ACKNOWLEDGMENTS}

We thank the anonymous reviewer for helpful comments and an interesting discussion. This work has made use of data from the European Space Agency (ESA) mission Gaia (https://www.cosmos.esa.int/gaia) processed by the Gaia Data Processing and Analysis Consortium (DPAC, https://www.cosmos.esa.int/web/gaia/dpac/consortium). Funding for the DPAC has been provided by national institutions, in particular, the institutions participating in the Gaia Multilateral Agreement.

\section*{FUNDING}

This study was conducted under the state assignment of Lomonosov Moscow State University. E.~N.~Podzolkova is a scholarship holder of the Foundation for the Advancement of Theoretical Physics and Mathematics BASIS (Grant No. 21-2-2-44-1).

\section*{CONFLICT OF INTEREST}

The authors of this work declare that they have no conflict of interest.

\end{document}